\documentclass[10pt]{article}

\usepackage{amsmath}
\usepackage{amssymb}

\usepackage{graphicx}

\usepackage{pdfpages}
\usepackage{cite}

\usepackage{color} 

\usepackage[labelfont=bf,labelsep=period,justification=raggedright]{caption}

\makeatletter
\renewcommand{\@biblabel}[1]{\quad#1.}
\makeatother

\date{}

\begin{document}

\begin{flushleft}
{\Large
\textbf{Bistability: Requirements on Cell-Volume, Protein Diffusion, and Thermodynamics}
}
\\
Robert G. Endres$^{*}$
\\
Department of Life Sciences \& Centre for Integrative Systems Biology and 
Bioinformatics, London, United Kingdom
\\
* r.endres@imperial.ac.uk
\end{flushleft}

\section*{Abstract}
Bistability is considered wide-spread among 
bacteria and eukaryotic cells, useful e.g. for enzyme induction, bet hedging, 
and epigenetic switching. However, this phenomenon has mostly been described 
with deterministic dynamic or well-mixed stochastic models. Here, we map 
known biological bistable systems onto the well-characterized biochemical 
Schl\"ogl model, using analytical calculations and stochastic spatio-temporal 
simulations. In addition to network architecture and   
strong thermodynamic driving away from equilibrium, 
we show that bistability requires fine-tuning towards small cell volumes 
(or compartments) and fast protein diffusion (well mixing). Bistability is 
thus fragile and hence may be restricted to small bacteria and eukaryotic 
nuclei, with switching triggered by volume changes during the cell cycle. 
For large volumes, single cells generally loose their ability for bistable 
switching and instead undergo a first-order phase transition.
  

\section*{Introduction}
Isogenetic cell populations show remarkable heterogeneity due to
unavoidable molecular noise - bacteria are either induced or
uninduced to produce enzymes for utilizing a particular sugar \cite{ozbudak04},
or enter cellular programs such as competence during starvation in
a reversible, switch-like manner \cite{suel06}. In higher organisms examples
of bistability are maturation in developing oocytes in {\it Xenopus} frog
embryos \cite{xiong03}, Hedgehog signaling in stem cells \cite{lai04}, and
phosphorylation-dephosphorylation cycles, e.g. as occurring in
mitogen-activated protein kinase (MAPK) cascades \cite{markevich04}. Bistable
pathway designs have also been explored in synthetic biology 
\cite{angeli04,pomerening03,kramer05,shimizu11,gardner00}. 
In analogy with physical bistable systems such as ferromagnets,
biological cellular systems can indeed exhibit hysteresis,
indicative of a system's memory of past conditions 
\cite{ozbudak04,angeli04,pomerening03,kramer05}.
Functional advantages of bistability include bet-hedging strategies,
decision-making, specialization, and mechanisms for epigenetic
inheritance, all increasing the species' fitness \cite{veening08,ferrell12}. 
However, these phenomena have mostly been described with deterministic
dynamic models or well-mixed stochastic models. It is unclear
if bistability predicted by the deterministic model always
corresponds to a bimodal probability distribution in the
stochastic approach \cite{vellela09}. Furthermore, the influence of 
slow protein diffusion and localization inside the cytoplasm (bacteria) 
or nucleus (eukaryotes) is often neglected. Whether
bistability is robust to such perturbations is unclear.

The question of the role of reaction volume in well-mixed bistable 
chemical reactions has a long history, e.g. \cite{schloegl72,vellela09,ge09,zuk12,nicola12}.
Particularly noteworthy, Keizer's paradox says that microscopic and macroscopic descriptions can 
yield different predictions \cite{vellela09}. In the macroscopic description the steady-state 
($t\rightarrow\infty$) is considered after taking the infinite volume limit ($V\rightarrow\infty$), 
while in the microscopic description the opposite order of limits is taken. Since the orders are not always 
interchangeable \cite{kurtz71,kurtz72,luo84}, unexpected results can occur. For instance, 
in the logistic growth equation species extinction occurs in the microscopic 
description, while the macroscopic description predicts a stable finite steady-state 
population \cite{vellela07}. As a consequence, in bistable systems it is generally not possible 
to derive Fokker-Planck or Langevin equations that produces a behavior in 
accordance with the master equation \cite{hanggi84,vellela09}. Derived potentials determining 
the weights of the states are incorrect. More sophisticated approximations (or modified Fokker-Planck 
approaches) are required to capture rare large fluctuations \cite{hanggi84,dykman94},
which ultimately determine the switching between states. 
However, the biological implications of these issues on cell-fate decisions have been 
rather unexplored, with some exceptions \cite{mehta08,vardi13}.

Furthermore, two recent papers address the effects of diffusion on 
bistability and switching of states. 
Zuk {\it et al.} considered a one-dimensional (1D) and a hexagonal 2D lattice 
model \cite{zuk12}, while T\u{a}nase-Nicola and Lubensky considered an 1D 
$M$-compartment model with hopping between the $M$ compartments to represent diffusion 
\cite{nicola12}. Based on their results, when the system size is small such systems are 
effectively well-mixed and transitions are driven solely by stochastic fluctuations in line with the
well-mixed master equation. However, when the system is spatially extended 
the more stable state spreads out in space and overtakes the more unstable state by the
mechanism of traveling waves. 
Interestingly, in presence of diffusion the stability of
steady states in the extended system is determined by the deterministic (mean-field) 
potential, which also describes the speed of the traveling waves.
However, it is unclear if these results also hold in 3D, for small volumes
comparable to nuclei and cytoplasms in cells, and using more realistic particle-based 
approaches.

Living cells are open molecular systems, characterized by
chemical driving forces and free-energy dissipation \cite{schroedinger44,vonbert50}. 
Here, we map known biological bistable systems onto the well-characterized 
non-equilibrium biochemical Schl\"ogl model \cite{schloegl72} (recently reviewed 
in \cite{vellela09}), allowing us to obtain analytical results for the well-mixed 
case. For slow diffusion we use stochastic spatio-temporal simulations. 
In addition to network architecture and strong 
thermodynamic driving away from equilibrium, we show that
bistability requires fine-tuning towards small cell volumes (or
compartments) and fast protein diffusion (well mixing). Bistability
is thus fragile and hence may provide upper limits on cell or nuclear
sizes. For increasing volume, a separation of time scales occurs 
and switching does not only become infinitesimally (exponentially) rare but
the weights of the states shift as well. Although states do not disappear per se,
weights can disappear, leading effectively to monostability. 
Hence, single cells loose their ability for reversible bistable switching and
instead undergo a first-order phase transition similar to mesoscopic
physical systems. Strict cell and nuclear size control may provide a
protective molecular environment for bistability. Indeed, our analysis of
previously published time-lapse movies of bacteria indicates that volume changes during
cell growth and division may function as triggers for switching.

\section*{Results}

\subsection*{Mapping of bistable systems onto Schl\"ogl model}
Bistability is driven by high-energy fuel molecules such as ATP and sources of
precursor molecules \cite{qian05}. Here, we choose the self-activating gene,
whose protein product binds its own promoter region to
cooperatively activate its own transcription as a dimer (see Fig. 1A,
mRNA is not explicitly modeled here). In addition to ATP
required for charging synthetase with amino acids and tRNAs,
high-energy molecules involved are nucleotide triphosphates during
transcription and GTP during translation \cite{hopfield74}. Additionally, we consider the
phosphorylation-dephosphorylation cycle with the phosphorylation
reaction catalyzed by kinase $K$ and the dephosphorylation reaction
catalyzed by phosphatase (inhibitor) $I$ (Fig. 1B) \cite{qian05}. Also in this
case there is positive feedback from product $P_p$ to its production.
The mean-field equations, given by ordinary differential equations
(ODEs), describe the temporal dynamics of the average protein
concentrations, valid in the limit of large volume (and hence large
protein copy numbers). At steady state, when all time-derivatives
are zero, the equations produce a bistable bifurcation diagram for
suitable parameter regimes (see Fig. 1C for a schematic). The
control parameter ($x$-axis) is a parameter of the model, e.g. a rate
constant, and the output ($y$-axis) is the target protein or the
phosphorylated protein concentration, respectively.\\

\begin{figure}[t]
\includegraphics[width=14cm]{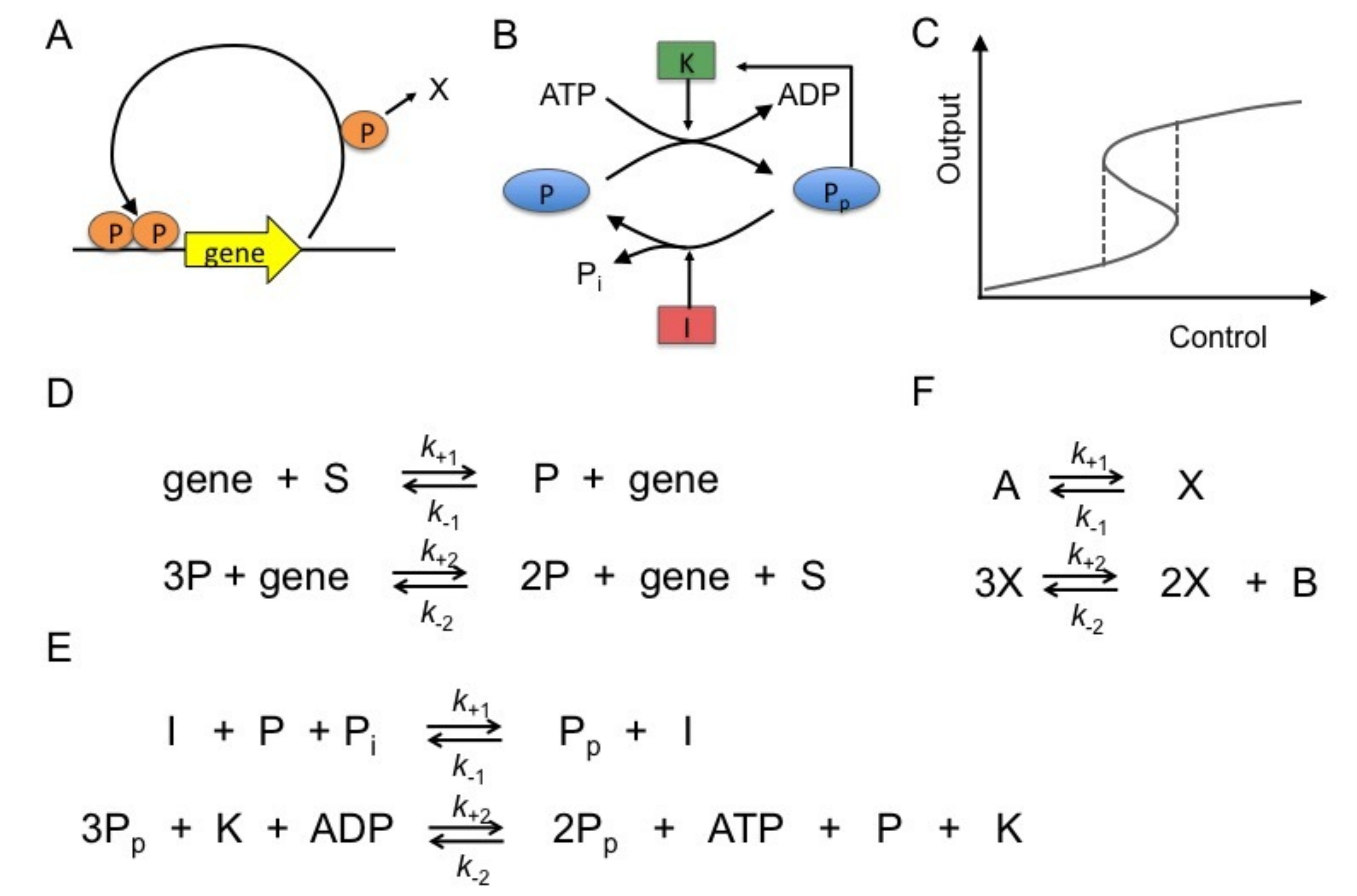}
\caption{{\bf Mapping of bistable systems onto Schl\"ogl model.}
(A) Self-activating gene with cooperativity. (B) Phosphorylation-dephosphorylation
cycle. (C) Schematic bifurcation diagram with bistable regime indicated by
vertical dashed lines. (D, E) Chemical reactions corresponding to (A) and
(B), respectively. (D) $S$ is substrate (nucleotides for mRNA and amino acids
for protein etc.) and $P$ is protein product. (E) Quantities $I$, $K$, $P$ ($P_p$), 
and $P_i$ are the inhibitor, kinase, (phosphorylated) protein, and inorganic phosphate,
respectively. (F) Chemical reactions of Schl\"ogl model with concentrations $A$
and $B$ adjustable parameters. For mapping reactions in (D) onto
reactions in (F) gene species needs to be absorbed into rate constants, and $S$ and
$P$ identified with $A/B$ and $X$, respectively. For mapping (E) onto (F) $I$, $K$,
ADP, and $P$ need to be absorbed into rate constants, and $P_p$ identified with
$X$, $P_i$ with $A$, and ATP with $B$.}
\label{Fig1}
\end{figure}

The chemical reactions for the self-activating gene and the
phosphorylation-dephosphorylation cycle with their rate constants
are shown in Figs. 1D and E, respectively. In Fig. 1D the first
equation indicates basal expression, while the second indicates
cooperative self-activation with $S$ the substrate (e.g. nucleotides for
mRNA and amino acids for protein production). In Fig. 1E, the
substrate is given by ATP, which is converted to ADP. $P_i$ is
inorganic phosphate produced during dephosphorylation, which
again is converted into ATP by the cell. Note, while the reverse
reactions from protein to substrate $S$ (Fig. 1D) and protein
phosphorylation by $I$ or protein dephosphorylation by $K$ are
extremely unlikely, they technically are nonzero and need to be
included for thermodynamic consistency. Importantly, the individual 
reactions can be mapped onto the well-characterized single-species Schl\"ogl
model, in which molecular concentrations $A$ and $B$ are fixed to
drive the reactions out of equilibrium. 

The mapping is justified based on the one-to-one correspondence 
of the molecular reactions (see Fig. 1F). For this, however, to work the biological examples 
would need to be implemented by mass-action kinetics instead of more realistic       
enzyme-driven kinetics. For instance, the self-activating gene
might be implemented by $dp/dt=a+bp^2/(K^2+p^2)-\tau^{-1}p$ to describe 
cooperative self-induction with Hill coefficient $2$, protein life time $\tau$,
and additional parameters $a, b$ and $K$.
While for the self-activating gene $p$-dependent production is to lowest order $\sim p^2$ similar to the 
Schl\"ogl model (with rate constant $k_{-2}$), its reverse rate is assumed to be zero as the forward rate is highly
driven by several enzymatic steps. In contrast, in the Schl\"ogl model the reverse rate is assumed to be 
non-zero (with rate constant $k_{+2}$). Similarly, while degradation in the Schl\"ogl model 
has a reverse rate (``accidental'' production from constituents via rate constant
$k_{+1}$) degradation in gene regulation is either implemented by active degradation 
or dilution during cell division, both of which have negligible reverse rates. As
a result, the macroscopic equation provided by the Schl\"ogl model is a third-order 
polynomial with rather large reverse reactions due to the absence of enzymatically driven
reactions.

\subsection*{Macroscopic perspective}
In the limit of large volume and hence
large molecule numbers, the Schl\"ogl model is described by ODE
\begin{equation}
\frac{dx}{dt}=-\underbrace{k_{+2}x^3}_{w_{+2}}+\underbrace{k_{-2}Bx^2}_{w_{-2}}
-\underbrace{k_{-1}x}_{w_{-1}}+\underbrace{k_{+1}A}_{w_{+1}}\label{Eq:ODE}
\end{equation}
with $x$ the molecular concentration. Once this limit is taken, time
can be sent to infinity. The resulting steady-state bifurcation
diagram is shown in Fig. 2A for standard parameters (see Materials
and Methods), with concentration $B$ chosen the control parameter.
Two saddle-node bifurcations (SNs) indicate the
creation/destruction of steady states, with a range of bistability
described by $\Delta B$ in between. However, the macroscopic perspective
makes no prediction about the relative stability of the two stable
steady states (black and blue curves with the unstable steady state
shown in red). In particular, do transitions simply become 
rarer with increasing volume so that the state attractors become increasingly 
deep but the relative weights of the states intact, or do the weights of the states 
change as well, leading effectively to loss of bistability? Furthermore,
is there a thermodynamic selection principle for the most stable
steady state? According to the second law of thermodynamics
entropy is maximized in a closed system at equilibrium. Does a
similar extremal principle hold for nonequilibrium steady states?
The rate of entropy production describes how much heat is
dissipated per time at steady state (and hence is a lower bound of how
much energy is consumed to maintain the state \cite{vellela09,gaspard04}): 
\begin{equation}
\frac{ds}{dt}=\sum_{i=1}^2(w_{+i}-w_{-i})\log\left(\frac{w_{+i}}{w_{-i}}\right)\geq0.
\end{equation}

\begin{figure}[t]
\includegraphics[width=14cm]{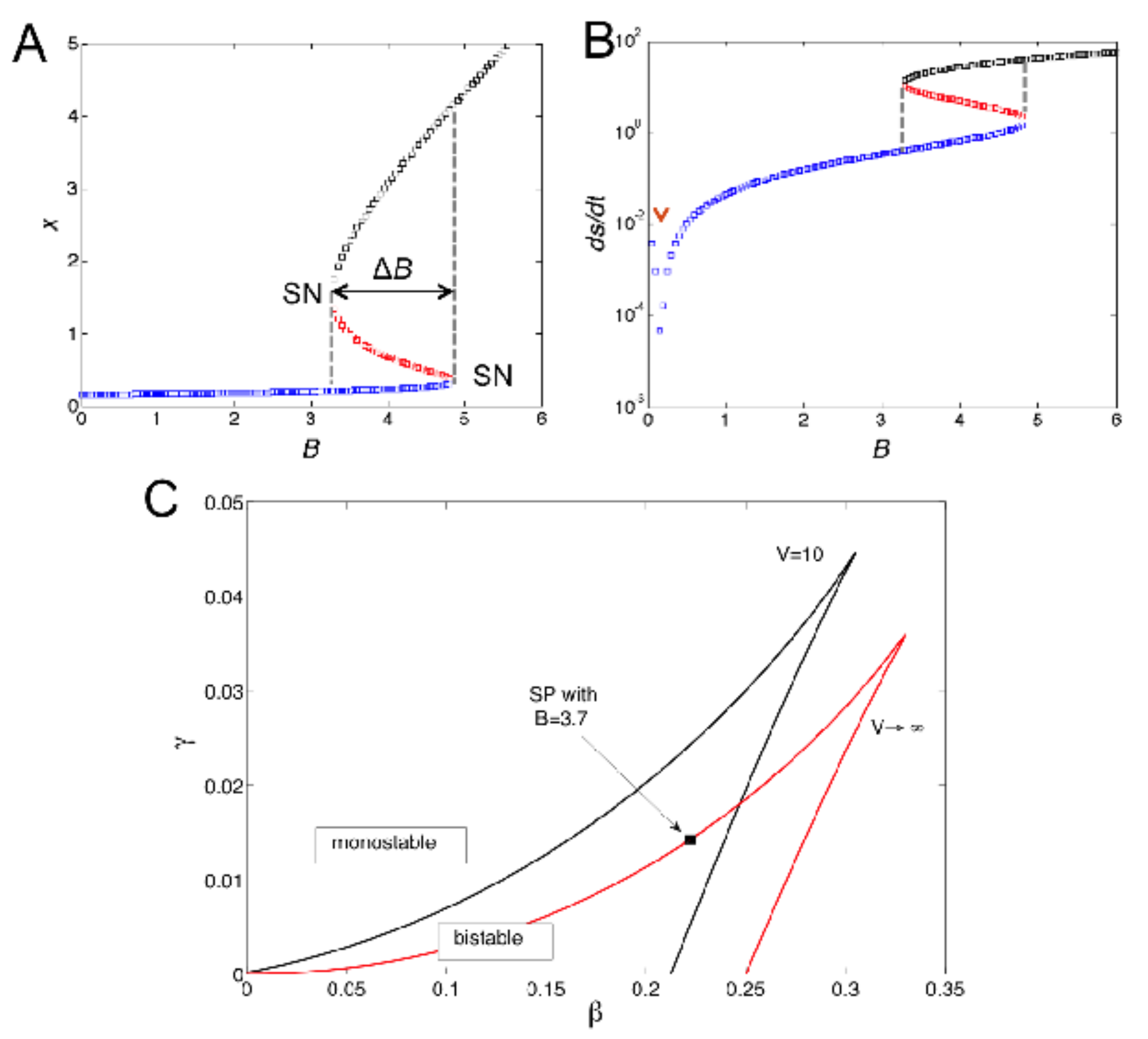}
\caption{{\bf Properties of macroscopic bistable system.}
(A) Bifurcation diagram $x(B)$ with the low stable steady state in blue, the unstable steady
state (saddle point) in red, and high stable steady state in black for standard
parameters defined in Materials and Methods. Black arrow indicates
bistable regime. (B) Corresponding entropy production rate as defined in
Eq. 2. (C) Phase diagram showing monostable states (only low or high
state) and bistable regions in $\beta$-$\gamma$ plane with 
combination parameters $\beta=k_{-1}k_{+2}/(k_{-2}B)^2$ and $\gamma=k_{+1}Ak_{+2}^2/(k_{-2}^3B)$.
Two phase diagrams correspond to $v=(k_{-2}B/k_{+2})V$ given by $37$ (exemplified by combination
$V=10, B=3.7$ and standard parameters; black lines) and $\infty$ (red lines). The latter
corresponds to the macroscopic mean-field model. SP indicates point 
$(\beta,\gamma)=(0.22,0.14)$ corresponding to standard parameters with $B=3.7$ 
(see S1 Text and \cite{ebeling79} for details).}
\label{Fig2}
\end{figure}

Intuitively, the entropy production is the net flux (difference
between forward and backward fluxes) times the difference in
chemical potential between products and educts (log term),
summed over all the reactions. Eq. 2 thus effectively describes how
quickly the maximum entropy state is reached, if left to equilibrate.
Prigogine and co-workers argued for a minimal rate of entropy
production, at least near equilibrium \cite{glansdorff71}, 
while others argued for
maximal rate of entropy production \cite{sawada81,dewar05}. 
Fig. 2B shows the
macroscopic rate of entropy production. The red arrow indicates
equilibrium at which the rate is zero. Notably the high state always
has the higher entropy production. This can easily be understood
\cite{andresen84} since overall in the Schl\"ogl model $A$ is converted to $B$ 
(and vice versa, see Fig. 1F). Hence, 
$ds/dt=\Delta G/T\cdot dA/dt~x$ with
$\Delta G$ the change in free energy for the overall reaction, $T$ the
temperature of the bath and $dA/dt$ a linear function of $x$. As a result,
if minimal entropy production is the rule, then the low state should
be selected, while maximal entropy production would dictate that
the high state is more stable. As eigenvalues of the 
Jacobian only contain information about local stability of a fixed point 
and not about global stability across multiple fixed points 
\cite{nicolis77,nicola12}, further discussion needs to be postponed until 
the next section. Fig. 2C summarizes the phase diagram 
(see S1 Text), showing monostable (low or high state only) and 
bistable regions in line with a cusp catastrophe. Limit $V\rightarrow\infty$,
shown by red lines, is relevant for the macroscopic description.

\subsection*{Microscopic well-mixed perspective}
When first taking the long-time
limit for a fixed finite volume to obtain the steady-state
distribution and then increasing the volume, we obtain a very
different picture of bistability. Assuming a well-mixed
microenvironment and thus neglecting diffusion (illustrated in Fig.
3A), we can employ the one-step chemical master equation to
describe the probability distribution in time
\begin{equation}
\frac{d}{dt}p(X;t)=\sum_{i=\pm1}^{\pm 2}[W_{i}(X-\nu_i|X)p(X-\nu_i;t)
-W_{-i}(X|X-\nu_i)p(X;t)]
\end{equation}
with $X$ the molecule copy number and $W_{+1}(X|X+1)=k_{+1}AV, W_{-1}(X|X-
1)=k_{-1}X, W_{+2}(X|X-1)=k_{+2}X(X-1)(X-2)/V^2$, and $W_{-2}(X|X+1)=k_{-2}BX(X-1)/V$ 
the volume-dependent transition rates. In Eq. 3, the sum is
over both the forward ($i=+1,+2$) and backward ($i=-1,-2$) reactions
with $\nu_{\pm 1}=\pm 1$ and $\nu_{\pm 2}=-\nu_{\pm 1}$ \cite{gaspard04}.
Using this description, we first simulate the master equation using
the Gillespie algorithm \cite{gillespie77} and confirm stochastic switching
between low and high stable states (Fig. 3B). However, at steady
state setting $dp/dt=0$, the probability distribution can analytically be
derived using a recursive relation leading to $p(x)=N(x)\exp[-V\Phi(x)]$
with potential \cite{hanggi84}
\begin{eqnarray}
\Phi(x)&=&x(\ln x-1)+x\ln\left(\frac{k_{+2}x^2+k_{-1}}{k_{-2}Bx^2+k_{+1}A}\right)\nonumber\\
&&+2\sqrt{\frac{k_{-1}}{k_{+2}}}\arctan\left(\sqrt{\frac{k_{+2}}{k_{-1}}}x\right)-
2\sqrt{\frac{k_{+1}A}{k_{-2}B}}\arctan\left(\sqrt{\frac{k_{-2}B}{k_{+1}A}}x\right)
\end{eqnarray}
and volume-independent prefactor
\begin{equation}
N(x)=\frac{k_{+2}x^2+k_{-1}}{Z\sqrt{x}[x^2+k_{+1}A/(k_{-2}B)]},
\end{equation}
where $Z$ is a normalization constant, $x=X/V$ and $p(x)=Vp(X)$ (see S1 Text 
for details). By construction, the stochastic potential also has
minima (a maximum) at the stable (unstable) deterministic steady
states (state) with
\begin{equation}
\frac{d\Phi}{dx}=-\ln\left(\frac{w_{+1}+w_{-2}}{w_{-1}+w_{+2}}\right)
\end{equation}
equal to zero at steady state ($w_{+1}+w_{-2}=w_{-1}+w_{+2}$) and $d^2\Phi/dx^2$ 
having the correct sign (see S1 Text for details and S1 Fig. for a plot of 
$\Phi(x)$). Fig. 3C shows indeed that the resulting distribution 
of $x$ has the expected bimodality.\\

\begin{figure}[t]
\includegraphics[width=14cm]{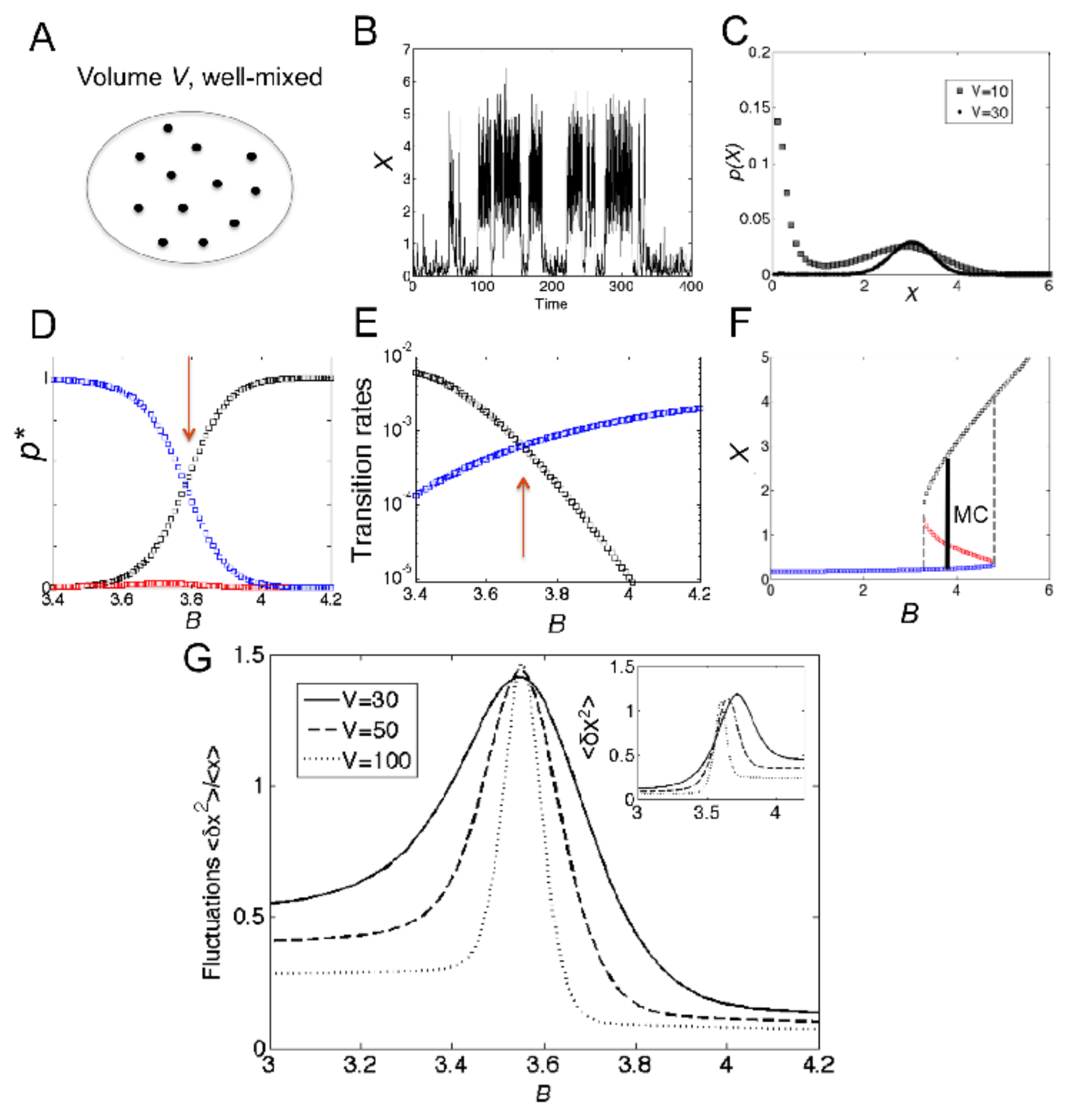}
\caption{{\bf Well-mixed bistable system.} (A) Schematic of well-mixed
system with volume $V$ (diffusion constant $D$ is infinitely large). (B)
Exemplar time trace for $x=X/V$ from Gillespie algorithm for standard
parameters with $V=10$ and $B=4.0$. (C) Exact probability distribution $p(x)$ at
steady state from master Eq. 3 for $V=10$ (dark symbols) and $30$ (light
symbols) with $B=4.0$. (D) Values of $p(x)$ evaluated at three steady states for
different values of $B$. (E) Transition rates from a modified Fokker-Planck
approximation valid for large $V$ (first-mean passage time; see S1 Text for details).
Red arrows indicate exchange of stability. (F) Maxwell-like construction
(MC), indicating coexistence between two phases (low and high states) 
at $B\sim 3.7$, defined by equal transition rates in (E). At this
critical value of $B$ a first-order phase transition occurs (see Text 
S1 for an analytical derivation based on simpler potential).                
(G) Relative strength of fluctuations (standard deviation over mean) as a
function of $B$ for $V=30$ (solid line), $50$ (dashed line), and $100$ (dotted line).
(Inset) Unnormalized variances.}
\label{Fig3}
\end{figure}

We are now in a position to address the relative stability of the
steady states, in particular of the two stable states. Fig. 3D shows
the probabilities evaluated at the deterministic steady states,
indicating a crossing of the stable states (exchange of stability) with
the probability of the unstable (metastable) state consistently below
the probabilities of the stable states. A more precise picture emerges when 
plotting the transition rates for switching between the stable steady
states in Fig. 3E \cite{hanggi84}, showing coexistence of 
the two stable states at $B\sim 3.7$. As derived in S1 Text, the rates depend
exponentially on the volume (as expected). However, due to 
normalization and the volume-independence of the prefactor, 
the more stable of the two becomes increasingly selected for larger and 
larger volumes, leading effectively to monostability. Fig. 3F shows that a 
Maxwell-type construction (MC) is required to establish the point of 
stability exchange, well known from the classical Van der Waals gas 
(see S1 Text for details) \cite{schloegl72,vellela09,ge09}. 
Since  the two states have different entropy productions (Fig. 2B, which can 
also be confirmed by calculating the microscopic entropy production
defined in S1 Text), we obtain a discontinuity at this point, indicative of a
first-order phase transition. Fig. 3G shows indeed a sharpening of
the molecular fluctuations at the critical point for increasing
volume. Hence, mesoscopic cells can loose their ability for bistability
({\it i.e.} a bimodal distribution) with increasing volume. 

The strong volume-dependent of bistability can also be
seen in the phase diagram in Fig. 2C (see \cite{ebeling79}). 
For small volumes (black lines) the region of bistability can significantly 
deviate from the corresponding region in the macroscopic limit (red lines).
For instance, a point in parameter space with strong bistability in the
microscopic system ($B=3.7$ for $V=10$) is borderline bistable in the 
macroscopic limit (cf. Fig. 3C for $B=4.0$). However, the phase diagram
does not contain information on the weights, and so shows a large
bistable region even in the macroscopic limit.

\subsection*{Microscopic perspective with diffusion}
Diffusion introduces inhomogeneous distributions of molecules, with diffusion
particularly slow in the crowded intracellular environment (Fig.
4A). For this purpose we turn to the stochastic {\it Smoldyn} simulation
package for implementing particle-based reaction-diffusion systems
in a box (Fig. 4B; see \cite{andrews10} and Materials and Methods for further
details). The third-order reaction (see Fig. 1F) needs to be
converted into two second-order reactions since no   
two events can exactly occur at the same time. (We call this model the 
generalized Schl\"ogl model.) This conversion requires
introducing of a dimer species $X_2$ with additional rate constants $k_{+3}$
and $k_{-3}$ as illustrated in Fig. 4C. For $k_{+3}=k_{-3}$ the steady-state values 
remain unchanged in the macroscopic limit (see S1 Text).
For reasonable diffusion constants (see Materials and Methods for parameter 
values), we indeed observe stochastic switching, resulting in a bimodal distribution 
for species $X$ (Fig. 4D). We then compared {\it Smoldyn} 
simulations in detail with  Gillespie simulations 
of the generalized and conventional reaction systems, including convergence 
for rare states with increasing simulation time, as well as effects of diffusion and
dimerization reactions on bimodal distribution (see S1 Text and S2-S4 Fig.). From these tests we conclude 
that {\it Smoldyn} simulations of the generalized system accurately produce 
bistable behavior, allowing us to study the effects of diffusion and volume 
on bistability.\\

\begin{figure}[t]
\includegraphics[width=14cm]{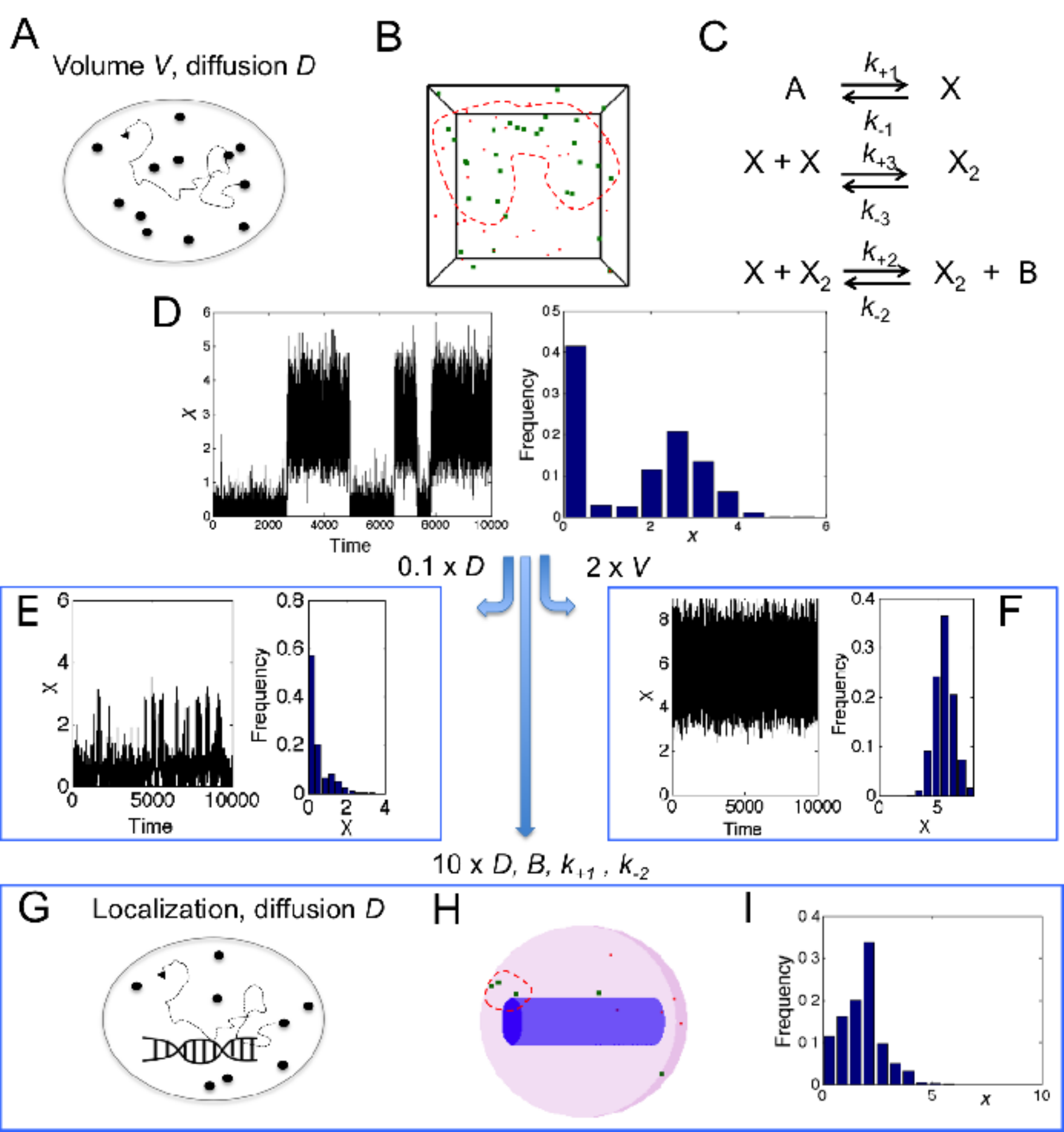}
\caption{{\bf Bistable system with diffusion.}  (A) Schematic of diffusing
molecules in volume $V$. (B) Snapshot of cubic reaction volume for
generalized Schl\"ogl model as simulated with {\it Smoldyn} software \cite{andrews10}. Shown
are monomers $X$ in red and dimers $X_2$ in green. Clustering is illustrated by
red dashed outline. (C) Chemical reactions of generalized Schl\"ogl model.
(D) Time trace (left) and histogram (right) of $x=X/V$ from simulation for
$D=3$ (for $X$) and 1 ($X_2$), $V=10$, $k_{+3}=k_{-3}=1$, and $B=3.7$. (E) and (F) Effects of
reduced (times $0.1$) diffusion constants (E) and increased (times $2$) volume
(F). In (E) $B=3.1$ to achieve comparable weights of low and high states. (G)
Schematic of localized transcription in self-activating gene pathway. (H)
Snapshot of spherical reaction volume with cylindrical DNA (purple) as
simulated with {\it Smoldyn}. Shown are monomers in red and dimers in green
with illustration of clustering by red dashed outline. (I) Histogram of
monomer concentration $x$ from simulation for $V=2.14$ and $V_\text{DNA}=1.51$, $D=30$
($X$) and $10$ ($X_2$), $k_{+1}=k_{+2}=50$ and $B=50$.}
\label{Fig4}
\end{figure}

Decreasing the diffusion constants of both molecular species by an
order of magnitude, suitable for macromolecular complexes or
membrane-bound proteins \cite{kalwarczyk12}, leads to strongly fluctuating
molecular concentrations (illustrated by the molecule cluster
enclosed by red dashed line in Fig. 4B) and reduced molecule numbers in the
high state (Fig. 4E).  When instead increasing
the reaction volume by just a factor $2$, the high state is strongly
induced (Fig. 4F). This result resembles the destruction of
bistability observed in the macroscopic limit (Fig. 3D-F).

In Fig. 4A-F the molecules are able to react anywhere in the
reaction volume. However, in cells, e.g. for a self-activating gene,
transcription occurs localized at the DNA molecule (Fig. 4G). To
investigate the effect of localization on bistability we use a
spherical cellular compartment (representing e.g. a bacterial cell or
a eukaryotic nucleus) in which we introduce a small cylinder to
represent the DNA molecule. The production can only occur in this
cylinder (Fig. 4H). In contrast, degradation can occur anywhere in
the cellular compartment. Fig. 4I shows that bistability is destroyed
with localized production, even for drastically increased production
rates and diffusion constants, which would easily produce
bistability under well-mixed circumstances. The broad distribution
in Fig. 4I may thus be caused by strong local fluctuations in
molecule number (illustrated by molecule cluster enclosed by red
dashed line in Fig. 4H). Note that the appearance of DNA as a
single copy is markedly different from the conventional or
generalized Schl\"ogl model, in which the molecule numbers scale
with volume. Next, we will explore the reasons for the breakdown
of bistability with inhomogeneity.

Fig. 5 shows a systematic exploration of bistability from diffusive
{\it Smoldyn} simulations, conducted similar to Fig. 4D.
Fig. 5A shows little evidence of bistability with the system either in the 
low or high state. Diffusion causes strong fluctuations in molecule numbers 
(and hence clustering) as as demonstrated by the radial pair-correlation function
$g(r)$ in Fig. 5B (not to be confused with the spike in fluctuations
at the critical point of the well-mixed system in Fig. 3G). For small molecule-molecule 
distances $r$, we obtain $g(r)\gg1$, which is the larger the slower diffusion. 
In contrast, random distributions of molecules do not show clustering.
While the next section investigates the role of such fluctuations in the loss of 
bistability, our findings are summarized in Fig. 5C, which shows the bistable range $\Delta B$ 
for the well-mixed case and the inhomogeneous case with finite diffusion constants. 
Here the system is considered bistable if simulations started from low and
high states exhibit at least one reversible switch within the simulation time 
(see figure caption for additional details).
The narrow range, especially for finite diffusion constants, suggests that
bistability is a fragile property, which needs protection. Indeed,
bacterial cell volume and nuclear volume in eukaryotic cells are
tightly regulated (e.g. nuclear volume does not simply scale with
DNA content \cite{marshall12}). When converting to physical units, our
predicted bistability regions fall nicely into experimentally
observed cell volumes (shaded areas in Fig. 5C). Importantly, as
volume varies during cell growth and division, such changes in
volume may function as a pacemaker or trigger for phenotypic
switching.\\

\begin{figure}[t]
\includegraphics[width=14cm]{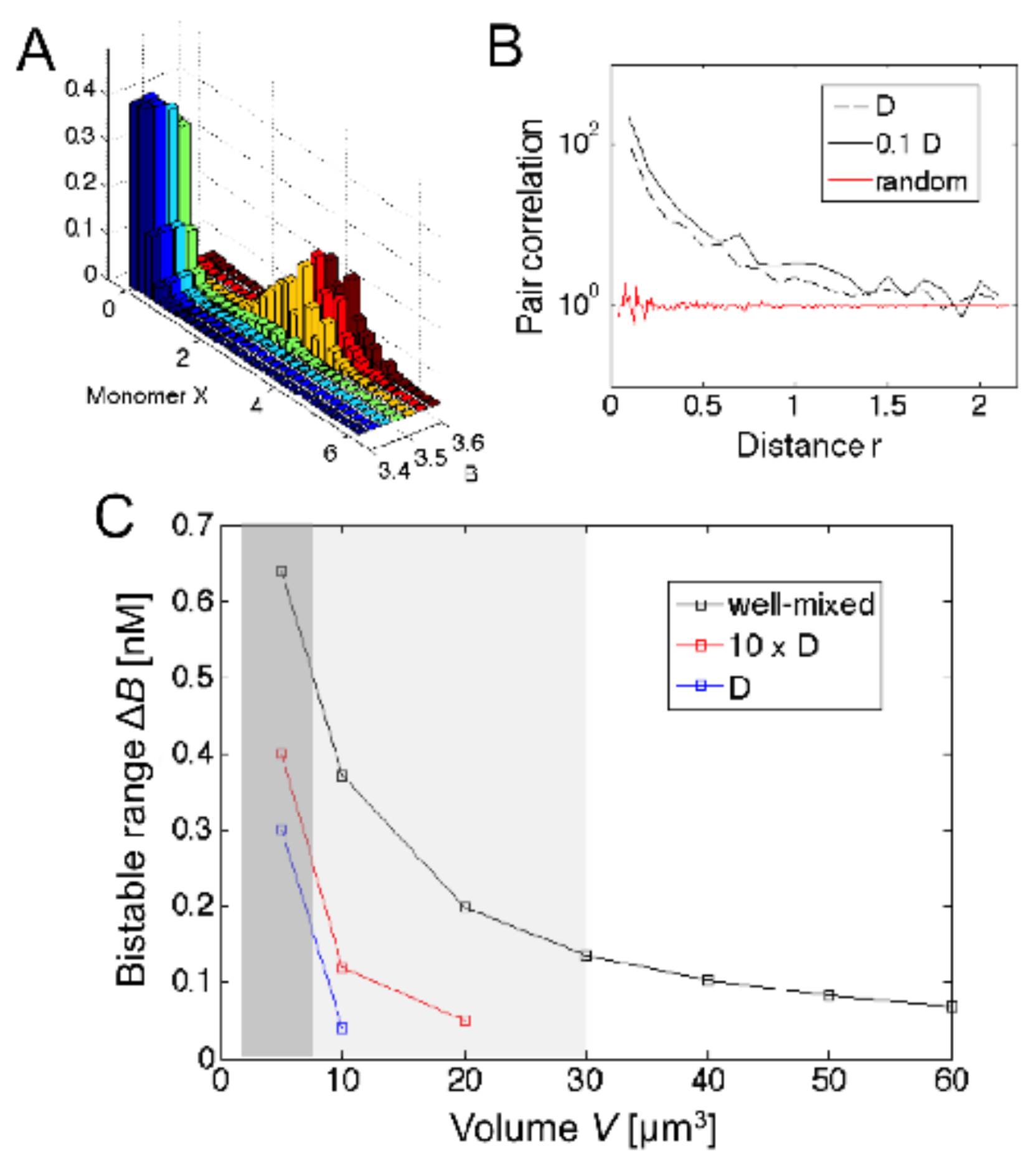}
\caption{{\bf Fragility of bistability.} (A) Histograms of monomer
concentration $x$ as a function of control parameter $B$ (from $3.4$ to $3.9$ in steps
of $0.1$) with other parameters chosen as in Fig. 4D. (B) Radial pair-correlation
function for $D=3$ ($X$) and $1$ ($X_2$) (dashed black line) and $D=0.3$
($X$) and $0.1$ ($X_2$) (solid black line) compared with random distribution (red
line; see Material and Methods for details). (C) Range of $B$ values with
visible bimodal distribution from master equation (well-mixed, black line)
and (inhomogeneous) {\it Smoldyn} simulations for $D=3$ ($X$) and $1$ ($X_2$) (blue
line) and $D=30$ ($X$) and $10$ ($X_2$) (red line) in units of $\mu m^2/s$, the latter being
typical protein diffusion constants in the cytoplasm. System was classified
bistable when 10,000-long simulations (see Materials and Methods) started in low and high 
states showed at least one reversible switch. Note that parameters
are converted to physical units here (see Materials and Methods for details).
Hence, a 10,000-long-simulation corresponds to a duration of 2.78h,
which is a very conservative estimate of cell-division times in bacteria and yeast.
Shaded areas indicate bacterial (dark) and eukaryotic nuclear (light)
volumes for comparison.}
\label{Fig5}
\end{figure}

\subsection*{Mechanism of bistability reduction by diffusion}

Increasing the volume shifts the weights of the states leading to
an effective loss of bistability (although the minima of the stochastic potential 
coincide with the deterministic model for sufficiently large volumes). 
How does slow diffusion affect bistability? There are two main potential reasons for the reduction of
bistability with diffusion: (1) Diffusion may increase the barrier of switching 
so that bistability is harder to achieve or observe, both in simulations and
experiments, or (2) diffusion may destabilize one of the stable states. In these
mechanisms local fluctuations in molecule numbers may play a role as well,
e.g. by introducing damaging heterogeneity or by nucleating traveling waves so that 
the more stable state can spread effectively and surpass the unstable state.
 
To rule out (1) longer and longer simulations can be conducted 
to guarantee convergence. S2 Fig. shows indeed that simulations are well converged,
even for weakly populated states. This shows that diffusion does not significantly change
the barrier height. To investigate (2) we use the method from \cite{kaizu14}
to renormalize the second-order rate constants of the generalized Schl\"ogl model 
($k_{\pm 2}, k_{\pm 3}$; cf. Fig. 4C) by diffusion (see Materials and Methods). This allows 
us to effectively include diffusion in the well-mixed model without having to conduct particle-based 
simulations. Fig. 6A shows that histograms from Gillespie simulations of the well-mixed 
Schl\"ogl model with renormalized reactions match well results from {\it Smoldyn} simulations 
(small Kullback-Leibler divergence). In contrast, Gillespie simulations without
renormalized reactions do not match well. In particular, without renormalization the switch to the high
state occurs at smaller $B$ values. Thus, achieving bistability is easier without diffusion 
as it requires less thermodynamic driving. These results are summarized in Fig. 6B by the
bifurcation diagram of the macroscopic model of the conventional Schl\"ogl model 
(Eq. \ref{Eq:ODE}) with renormalized rate constants $k_{\pm 2}$ (note 
$k_{\pm 3}$ do not affect the steady-state probability distribution as they are equal, 
see S1 Text). Specifically, this figure shows a significant delay in achieving bistability
with increasing $B$ value. Completely removing first and second terms in macroscopic Eq. \ref{Eq:ODE}  
leads to the complete collapse of bistability and a truly monostable state around 
$x=k_{+1}A/k_{-1}\approx 0.17$ in line with simulations (see S5 Fig.).\\

\begin{figure}[t]
\includegraphics[width=12cm]{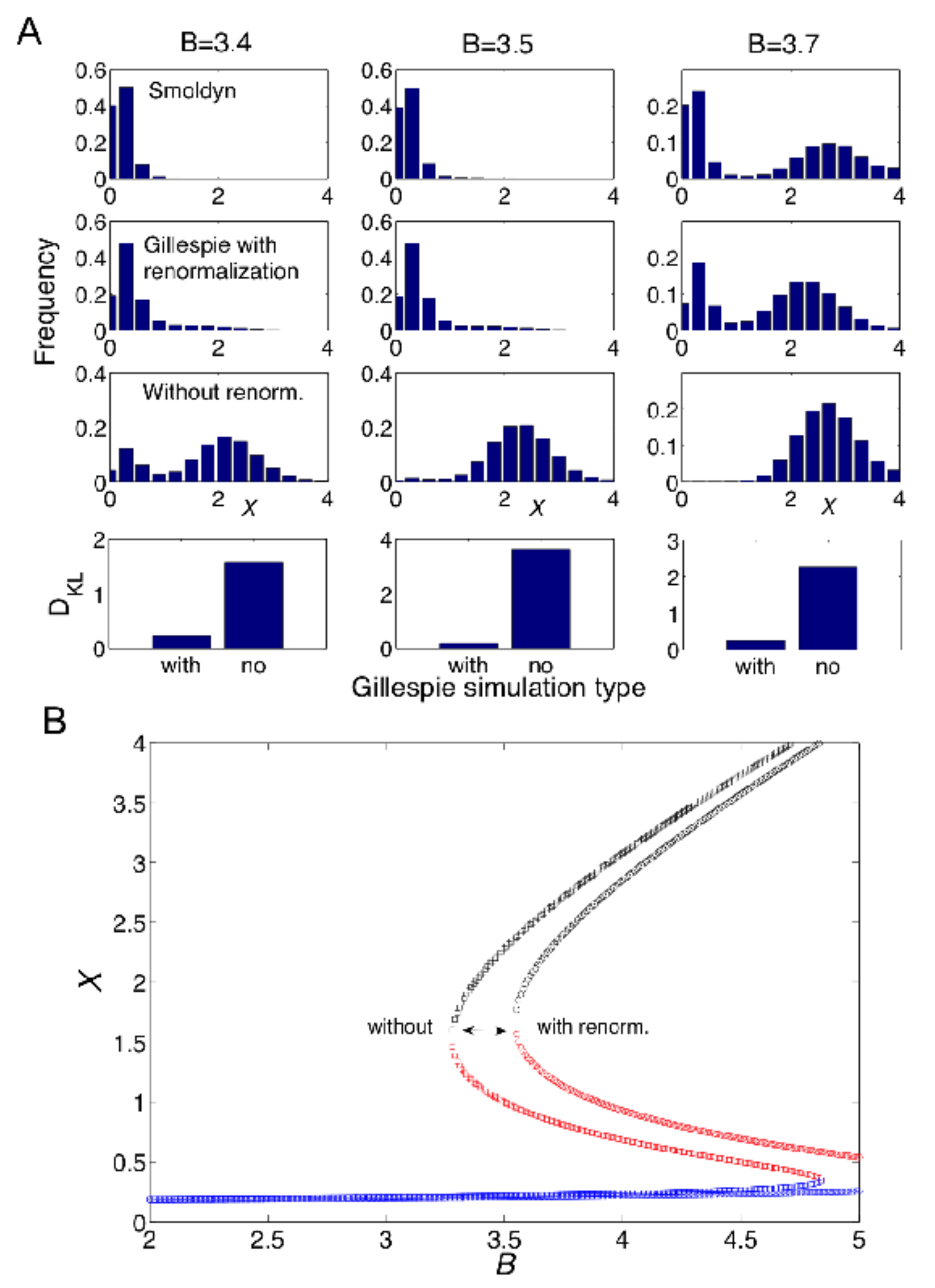}
\caption{{\bf Diffusion can be included by renormalization of
second-order rate constants.} 
(A) Histograms of monomer frequency for different $B$ values from simulations with {\it Smoldyn} 
software \cite{andrews10} (first row) and Gillespie simulations of the generalized Schl\"ogl model 
with (second row) and without (third row) renormalized rate constants of second-order reactions. 
Standard parameters were used with volume $V=10$. The Kullback-Leiber divergence ($D_{KL}$) shows
the closer correspondence of the renormalized reactions than the normal reactions
with {\it Smoldyn} (forth row). For details on renormalization and calculation of $D_{KL}$, 
see Materials and Methods. (B) Corresponding macroscopic bifurcation diagram of deterministic 
ordinary-differential equation model using renormalized rate constants $k_{\pm2}$ to illustrate 
effect of diffusion. This shows that diffusion delays entry into bistable regime for increasing $B$.}
\label{Fig6}
\end{figure}

What role might the fluctuations observed in Figs. 4B and 5B play?
Following ideas from extended bistable spatial systems \cite{zuk12,nicola12}, 
fluctuations may nucleate traveling waves, which then spread by diffusion.
Although our main interest are small systems most relevant to cell biology, 
we extended the simulation box in one of the spatial directions (Fig. 7A).
Kymographs from simulations with standard parameters, run for different $B$ values, 
show the spreading of the more stable state when initially started in the unstable 
state. Near co-existence at $B\sim 3.7$, traveling waves exist which do not
change the state permanently, but ripple through the box. Although wave velocities
can technically be obtained from the slope in the kymographs they are highly
variable and hard to determine objectively due to small molecule numbers.\\

\begin{figure}[t]
\includegraphics[width=14cm]{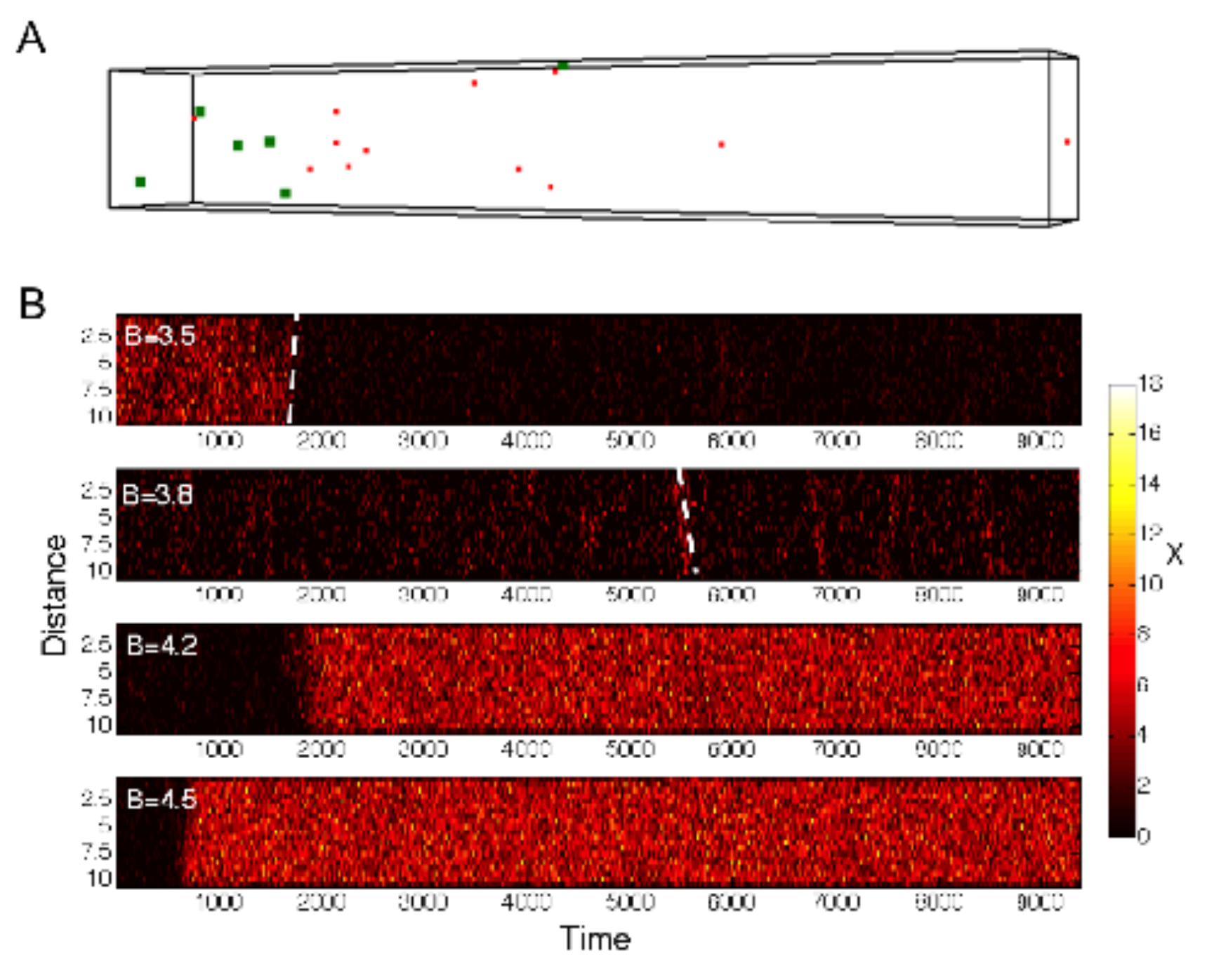}
\caption{{\bf Onset of traveling waves in spatially 
extended system.} (A) Snapshot of elongated reaction volume for generalized Schl\"ogl model 
as simulated with {\it Smoldyn} software \cite{andrews10}. Shown are monomers $X$ in red and 
dimers $X_2$ in green. (B) Kymographs of monomer numbers along major axis of simulation box
(distance) as a function of simulation time. For this purpose box was divided into 20 equal sized bins.
Parameter values: Standard parameters were chosen with volume of simulation box 
$V=10\cdot 1.5\cdot 1.5$, $B$ values as indicate in subpanels of (B), and other 
parameters as in Fig. 4D. Steepness of white dashed lines illustrates magnitude of 
wave velocity.}
\label{Fig7}
\end{figure}

Taken together, slow diffusion makes reaching bistability harder 
as second- (and higher-order) reactions are impaired - molecules have difficulties 
encountering each other to produce nonlinear behavior. Fluctuations may lead to 
traveling waves in more extended spatial systems, which provides a mechanism 
for the more stable state to overtake the less stable state.

\subsection*{Experimental prediction on switching with cell-volume changes}

Our models make strong predictions on the effect of cell volume.
One obvious prediction is that when a system is tuned towards the
bistable regime (which becomes harder and harder to achieve for
increasing volume), switching between the two states becomes
increasingly rare. This is the well-known transition from the
stochastic to the macroscopic, deterministic limit, and was recently
demonstrated using time-lapse microscopy. In budding yeast
({\it Saccharomyces cerevisiae}) the variability in the G1 phase (i.e. the
time from division to budding) is reduced with increased ploidy
(copies of chromosomes) \cite{ditalia07}. Similarly, the switching time for
turning the Pho starvation program off under reversal of phosphate
limitation is reduced with ploidy \cite{vardi13}. As the volume also scales
with ploidy, the protein concentrations stay approximately constant,
thus reducing cell intrinsic noise and hence stochastic transitions between
cellular states.

Our results, however, are more specific. They suggest increased 
unidirectional switching and hence monostability and decision-making in growing 
and dividing cells. In fact, growth towards cell division leads to a volume increase 
by a factor two, which may cause cells to select a steady state (see Fig. 4F).
Hence, for cellular parameters below the critical point, the low state
will be selected, while for cellular parameters above the critical
point, the high, induced state will be selected. Fig. 8A,B show the
expression of bistable reporters as a function of time, specifically
LacY in {\it E. coli} \cite{choi08} and ComK (or ComG) in {\it B. subtilis} 
\cite{suel06}. The latter is indicator for competence (while competence 
is strictly speaking an excitatory pathway, the core module of ComK is
bistable with an exit mechanism based on ComS \cite{suel06}). \\

\begin{figure}[t]
\includegraphics[width=14cm]{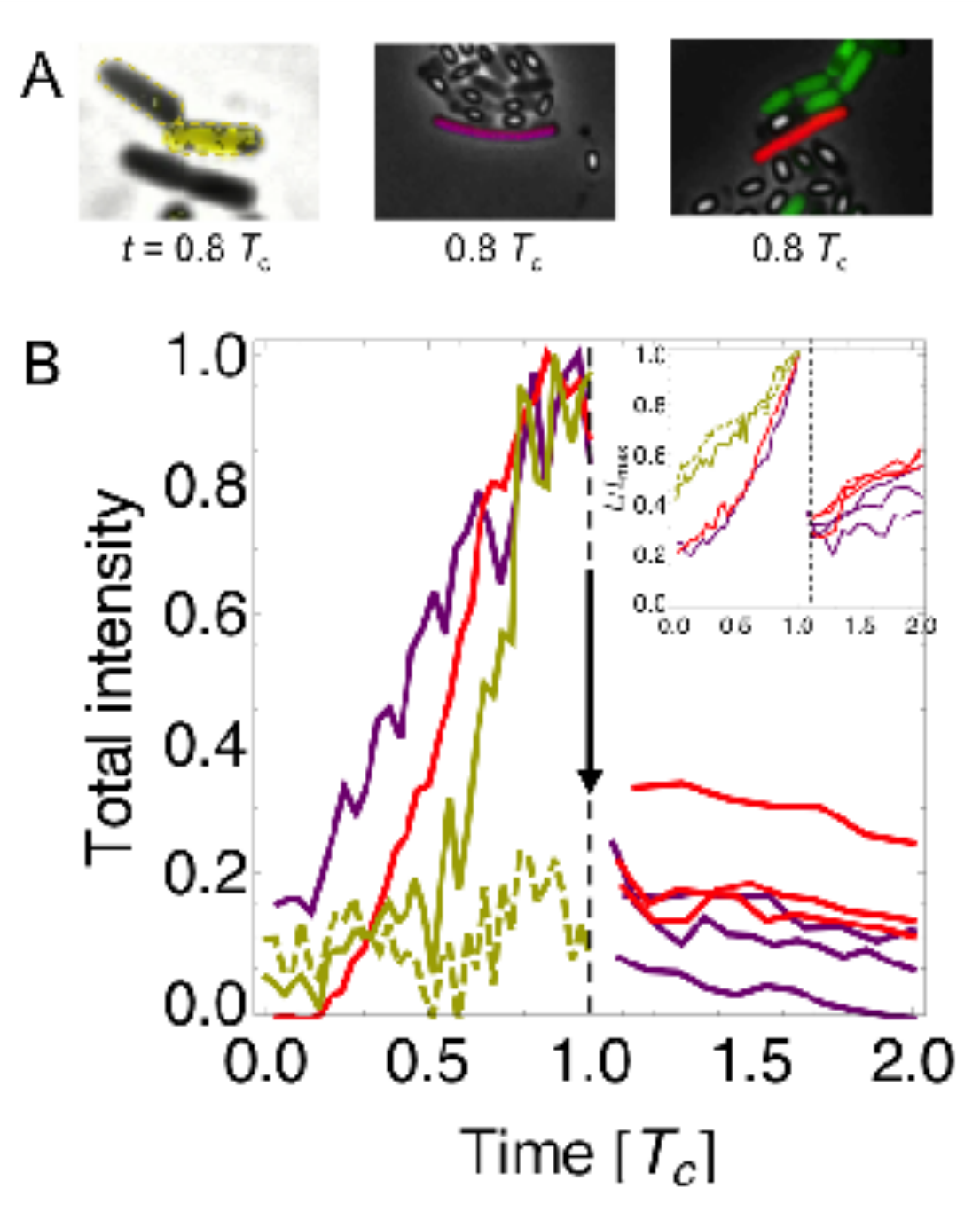}
\caption{{\bf Switching may be triggered by cell-volume changes.} 
(A) Snapshots from time-lapse fluorescence microscopy: (left) lacY-gfp of {\it E.
coli} in yellow \cite{choi08}, (middle) PcomK-cfp of {\it B. subtilis} in purple, and (right)
PcomG-cfp of {\it B. subtilis} in red \cite{suel06} with time in units of cell-cycle time
$T_c$. (B) Total fluorescence intensities inside cell contours normalized to the
maximal observed total intensity of a cell (see Materials and Methods for
details) with color-coding same as in panel (A). Two yellow daughter cells
are shown by solid and dashed lines. Note also the appearance of multiple
red and purple daughter cells right after cell division in competence. (Inset)
Normalized cell lengths over time in units of maximal cell length $L_\text{max}$. S6 Fig.
shows same for intensity density, {\it i.e.} total intensity divided by cell
area.}
\label{Fig8}
\end{figure}

Fig. 8B shows that switching to the high state appears during growth, while
switching to the low state occurs immediately after cell division
when the volume has shrunk suddenly to half its value.   
Note it is unlikely that the rise (drop) in fluorescence intensity
is simply due to switching induced by gene duplication (halving) as the
concentrations stay roughly constant due to accompanying volume
changes. Furthermore, the ratio of chromosome-replication and cell-division 
times is known to be about 2:1 \cite{botello98,zaritsky11}. Since, by visual inspection, 
cell division takes about 10\% of the cell-cycle time ($T_c$) in the time-lapse movies, 
chromosome replication takes about 20\% of it. This duration is short compared 
to the rise-in-intensity phase, suggesting a different mechanism.
Switching is still stochastic as shown by the two yellow daughter cells - one
induces the lac operon, while the other does not. In support for our
proposed scenario, spontaneous switching is extremely rare (for the lac operon
estimated to be around $0.004$ per cell cycle in presence of $40\mu$M
inducer TMG \cite{choi10}). Hence, volume changes during cell growth and 
division may instead be the main drivers, like a pacemaker, for switching.

\section*{Discussion}

We presented a nonequilibrium thermodynamic model of
bistability, relying on molecular stochasticity and chemical energy
for switching and decision-making. To cover a large class of
bistable systems, including self-activating genes with cooperativity
and phosphorylation-dephosphorylation cycles, we mapped
minimal models for these onto the well-characterized
nonequilibrium Schl\"ogl model. Bistability and its hallmark of
hysteresis are generic behaviors that are the same from one system
to the next regardless of details. Indeed, this property is shared with
ferro-magnets and mutually repressing genes (toggle switch)
\cite{gardner00,laughlin00}. Our approach is markedly different from recent
deterministic approaches to postulate multistability in signaling
cascades, which neglect the physical effect of cell volume and
molecular diffusion \cite{barik10}. Deterministic approaches 
often predict complex dynamics with multiple attractors. 
However, when the volume is sufficiently large, such behaviors can disappear. 
Not only does switching become increasingly rare, but also the weights 
shift and ultimately favor one of the states.
Hence, bacterial cells and eukaryotic nuclei, 
and cell compartments in general, may represent protectorates of complex bi- and multistable
behavior \cite{laughlin00}. In contrast, mesoscopic cells are ``boring'',
unable to display complex behavior.

Slow diffusion, caused by molecular crowding and localization, is a
killer of bistability and cells need to deal with this issue. 
This is because slow diffusion selectively penalizes second- and higher-order reactions
and hence nonlinearity. Consistent with our study, ultrasensitivity in MAPK cascades 
is destroyed for slow diffusion due to rebinding of enzymes to their substrate \cite{takahashi10},
stressing the fundamental importance of diffusion in theoretical
predictions of bistability. Phase domains and their movement are
well known from the Ginzburg-Landau equation for phase
transitions - this equation is in fact similar to the Schl\"ogl model
with diffusion (albeit in absence of stochastic effects). How can
cells cope with the negative effects of diffusion? While adjustment
of diffusion constants is difficult \cite{gregor05}, cells could use small
transcription factors to speed up diffusion. Up to about 110 kDa,
the mean diffusion coefficient falls close to the Einstein-Stokes
prediction for a viscous fluid \cite{gregor05}. This suggests that proteins up to
this size do not encounter significant diffusion barriers due to
macromolecular crowding or a meshwork of macromolecular
structures in the cytoplasm. Indeed, the repressor LacI of the {\it E. coli}
lac system, master regulator ComK of the {\it B. subtilis} competence
system, and transcription factor Gal80 of the gal system in budding
yeast are only $38.6$, $22.4$, and $48.3$ kDa large, and hence are
expected to have relatively large diffusion constants of at least 
$8\mu m^2/s$ (based on scaling relation in \cite{nenninger10}). 
Another option for the cell is to tune the viscosity of its cytoplasm below a   
glass-transition point where metabolism-driven active mixing produces superdiffusive
environments \cite{parry14}. Note, however, that cells need to protect 
themselves from sudden shrinkage of the cytoplasm or drastic increases 
in concentration by osmotic shock.

Due to small volumes and slow diffusion in cells, bistability can 
only occur in a narrow range of parameter space and thus may require fine-tuning
\cite{hermsen11} or a pacemaker. Consistent with this notion, our analysis of
time-lapse microscopy movies shows that volume changes during
the cell cycle may trigger switching events (Fig. 8). Such assistance
might be necessary since spontaneous switching can be extremely rare, likely
caused by rare bursts in gene expression \cite{choi08,choi10,boulineau13}. For
instance, the switching rate of lac system was estimated to be only
$0.004$ per cell cycle (in presence of $40\mu$M inducer TMG) \cite{choi10}, and
diauxic shifts take on average $2$ hours \cite{nenninger10} 
(see \cite{shimizu11,zong10,acar08} for
additional examples of slow switching much beyond the cell cycle
period). Taken together, switching may hence be more likely to occur via a
thresholding mechanism \cite{hermsen11} as implied by the Maxwell-type
construction. To clarify the details of the trigger mechanism,
more experimental investigation will be needed using
modern microfluidic designs for continuous imaging of cells over
very long times.

While some of the issues raised here have individually  
been discussed for the Schl\"ogl model before \cite{schloegl72,vellela09,ge09,zuk12,nicola12}, 
this has hardly been done in the context of biology. In particular, bistability and 
diffusion restrict volumes to the sizes of bacteria or eukaryotic nuclei, and volume 
changes during the cell cycle may be exploited to robustly change cellular states. 
Our particle-based simulations focus on small 3D volumes, most relevant to cellular
nuclei or cytoplasms, and hence are markedly different from recent studies
\cite{nicola12,zuk12}. The latter addressed role of volume and diffusion on bistability 
in extended 1 and 2D lattice models, respectively. Our loss of bistability
for slow diffusion is largely determined by renormalization of second-order
rate constants, which ultimately leads to a collapse of the system to the low state
for very small diffusion constants. When extending the system in one of the 
spatial dimensions, we observe the onset of traveling waves, which may ultimately 
determine switching rates for even larger systems. Specifically, for extended systems 
the wave velocity is determined by the deterministic (and not the stochastic) potential 
\cite{nicola12,zuk12}. Such quasi-1D extended states may biological be relevant in filamentous 
bacterial cells. Traveling waves are indeed observed in {\it Smoldyn} simulations of 
the Min-system, a small biochemical pathway which allows cells to determine their 
middle for accurate cell division \cite{hoffmann14}.

Bistability is fascinating due to its connection with nonequilibrium
physics, first-order phase transitions, and decision-making in cells.
Our results show that {\it volume shifts the weights} of the states relative to 
each other but not the steady-state values directly. Below a critical value the low state is 
selected, while above a critical value the high state is selected. In contrast, {\it diffusion 
shifts the steady-state values}. For sufficiently slow diffusion, only the low state survives.
While widely studied there are a number of open questions. One
pressing question is whether epigenetic information is inherited in the
spirit of Hopfield's content-addressable memory \cite{hopfield84}. Since the
seminal work by Novick and Weiner in $1957$, such inheritance
seems indeed to apply to the Lac system \cite{novick57}, while Fig. 8 shows that
the daughter cells do not generally inherit the competence state (this
might be due to the protease MecA \cite{suel06}). Furthermore, cell volume
changes and their effect on bistability also have many biomedical
implications. Examples include viruses such as bacteriophage
lambda \cite{zong10} and HIV \cite{singh09}, as well as pancreatic $\beta$ cells, responsible
for glucose sensing and insulin production. These cells undergo
large size changes, e.g. during pregnancy \cite{granot09}. Thermodynamics
may shed new light on their regulatory mechanisms.

\section*{Materials and Methods}

\subsection*{Schl\"ogl model}
In 1972 Schl\"ogl proposed two chemical reaction models for  
nonequilibrium phase transitions \cite{schloegl72}. One example shows a 
phase transition of first order, while another shows a phase transition of 
second order. When diffusion is included in the the first-order transition, 
coexistence of two phases in different spatial domains may occur. For spherical domains 
the coexistence indicates the onset of the transition similarly to the vapor 
pressure in droplets or bubbles. The volume dependence has been discussed 
early, e.g. in \cite{ebeling79}. The related Keizer's paradox has mostly been 
discussed in context of Schl\"ogl's first model, but also for the logistic growth 
equation \cite{vellela07}, showing its relevance to a wide sectrum of systems. 
In these systems, large rare fluctuations have severe consequences. Keizer's
paradox says that deterministic and stochastic approaches can lead to 
fundamentally different results. In particular, the deterministic model considers the 
infinite volume limit ($V\rightarrow\infty$) before considering the steady-state limit 
($t\rightarrow\infty$). Hence, no transitions between steady states are allowed. The
state of the system only depends on the initial condition, which seems unphysical. In
contrast, the stochastic model, by taking the steady-state limit first, can always settle 
in its lowest state, and thus is ultimately favored over the deterministic approach.
Gaspard \cite{gaspard04} and later Qian \cite{vellela09} took the perspective of
open chemical system, and analyzed the second Schl\"ogl model in terms of fluxes and 
entropy production. The Schl\"ogl model can be considered the simplest bistable
system but has not yet been verified or implemented experimentally by suitable 
chemical reactions.

\subsection*{Implementation and parameters} 
The chemical reactions of the Schl\"ogl model can be
found in Fig. 1F. Macroscopically (for an infinite volume) this
model can be described by ODE Eq. 1. For finite volume but
infinitely fast diffusion (well mixed case) the master equation (Eq.
3) can be used or Gillespie simulations \cite{gillespie77}. For the solution of the
master equation and a derivation of the transition rates see S1 Text. The
macroscopic (microscopic) entropy production formula is provided
in Eq. 2 (in S1 Text). The chemical reactions for the generalized Schl\"ogl
model are given in Fig. 4C with further details provided in S1 Text.

Stochastic spatio-temporal simulations with diffusion were
conducted with {\it Smoldyn} software version 2.28 as described in
\cite{andrews10}. Briefly, it is a particle-based, fixed time-step, space-continuous
stochastic algorithm for reaction-diffusion systems in various
geometries based on Smoluchowski reaction dynamics \cite{andrews04}. 
Both simulations and Smoluchowski theory only apply to reactions 
up to second order.  Given reaction rate constants, diffusion constants, and 
time step, {\it Smoldyn} determines reaction radii, {\it i.e.} binding and unbinding radii. 
The binding radii correspond to the encounter complex, formed by diffusion. Once formed, the
reaction occurs. Intrinsic rate constants are strictly constant, i.e. independent of diffusion, 
for low particle densities and activation-limited reactions (see manual for details). 
This is checked and confirmed by {\it Smoldyn} at the beginning of each
simulation. Under these conditions, the effects of rate constants and diffusion 
constants on bistability can independently be explored.

The renormalization of the second-order rate constants to 
effectively include diffusion into the well-mixed generalized Schl\"ogl model 
is done following \cite{kaizu14}. Using $k_D=4\pi\sigma D$ to describe
the encounter of two molecules by diffusion, the following rate constants are
obtained 
\begin{equation}
k_{\pm i}'=\frac{k_{\pm i}k_D}{k_{+i}+k_D}
\end{equation}
with $i=2,3$ (cf. Fig. 4C), $k_{\pm i}$ comparable to {\it Smoldyn}'s intrinsic rate constants, 
product of cross section $\sigma$, and average diffusion constant $D$ set to $0.5$.

Unitless parameters were generally used as given in \cite{gaspard04}, termed
standard parameters. These rate constants are $k_{+1}A=0.5, k_{-1}=3$, and
$k_{+2}=k_{-2}=1$. Only concentration $B$, diffusion constants, and 
volume $V$ were varied as indicated in figure captions. For {\it Smoldyn}
simulations we additionally used $k_{+3}=k_{-3}=1$ (Fig. 4D-F) and $V=2.14$,
$V_\text{DNA}=1.51, D=30$ ($X$) and $10$ ($X_2$), $k_1=k_2=50$ and $B=50$ 
(Fig. 4I). Simulation time was generally $10,000$ unless 
specified differently. To convert to units in Fig. 5C, we set length and 
time scales to respective $\mu m$ and $s$. We further express concentration 
in [nM], with $1$nM corresponding to $1$ ($1000$) molecules in a typical bacterial
(eukaryotic HeLa) cell, and volume as $V=\eta 10\mu m^3$ with $\eta$ 
a scaling number of order $1$. Rate constants then become $k_{+1}A=5/(6\eta)$ 
[nM/s], $k_{-1}=1$ [$s^{-1}$], $k_{+2}=B$ [$s^{-1}$], $k_{-2}=k_{+3}=6\eta/10$ 
[(nM s)${}^{-1}$], and $k_{-3}=1$ [$s^{-1}$]. For gene expression, $k_{+1}A$ 
corresponds to typical basal expression rates found in bacteria \cite{friedman06}.

\subsection*{Simulation analysis} The radial pair-correlation functions in 
Fig. 5B were calculated from 10 simulation snap shots of monomer $X$
positions $\mathbf{r}_i$ in a 3D cube of volume $V$, using
\begin{equation}
g(r)=\frac{V}{N(N-1)}\frac{1}{4\pi r^2 a}\sum_{i=1}^N\sum_{j\neq i=1}^N
I_{ij}(r-a<||r_j-r_i||\leq r)
\end{equation}
for $r\geq a$ with $a$ the mesh spacing, assuming periodic boundary
conditions. $I$ is either one if its argument is true or zero if false. $N$ is
the current monomer number. For plot $g(r)$ was then averaged over
snapshots for different $N$. Note since $N$ varies, $g(r)$ can
systematically deviate from $1$. As a control, random positions were
produced for which $g(r)\sim 1$ as expected.\\

\noindent The comparison of distributions in Fig. 6A is done with the 
Kullback-Leibler divergence, defined by
\begin{equation}
D_{KL}=\sum_i p_R(x_i)\ln\left[\frac{p_R(x_i)}{p_T(x_i)}\right],\label{Eq:DKL}
\end{equation} 
where $p_R(x_i)$ and $p_T(x_i)$ are the reference and test distributions, 
respectively. The sum in Eq. \ref{Eq:DKL} is over bins in 
$x$ space.

\subsection*{Image analysis} Fluorescence intensities of bacterial cells were
extracted from time-lapse movies of fluorescence microscopy ({\it E.
coli} from \cite{choi08} and {\it B. subtilis} from \cite{suel06}). The active-contour method
from \cite{smith10} as an {\it ImageJ} plugin was used to detect the cell
boundaries as ridges, and pixel intensities inside of the contours
were collected using a simple custom written {\it Mathematica} code.
The background intensities including the phase contrast intensities
were subtracted and only the intensities from the fluorescence
channels were plotted in Fig. 8B. Intensity density plotted in S6 Fig.
is calculated as total intensity of a cell divided by the cell
current focal area.


\section*{Acknowledgments}
We thankfully acknowledge Hong Qian for constructive comments
on the manuscript, and Nikola Ojkic for help with the image
analysis. Financial support was provided by Leverhulme-Trust
Grant N. RPG-181 and European Research Council Starting-Grant
N. 280492-PPHPI. This work was initiated at the workshop
“Information, probability and inference in systems biology” (IPISB
2013) in Edinburgh.

\section*{Supporting Information Legends}

\noindent{\bf S1 Text:} Supporting text with mathematical derivations and additional explanations.\\
\ \\
\noindent{\bf S1 Figure:} Comparison of deterministic and stochastic potentials. For concentration $x$ 
the deterministic potentials $\Psi(x)$ in green is calculated with Eq. 44 in S1 Text, while the stochastic potential 
$\Phi(x)$ in blue is calculated with main-text Eq. 4  (Eq.  19 of S1 Text). (A) Arrow indicates that for $B=3.0$ 
both potentials predict the low state as most stable. (B) Arrows indicate that around $B=3.4$ the 
deterministic potential predicts coexistence of the low and high states. (C) Arrows indicate that around
$B=3.7$ the stochastic potential predicts coexistence. (D) At $B=4.0$ both potentials
predict the high state as most stable.\\
\ \\
\noindent{\bf S2 Figure:} Bimodal distribution with rare high state (low weight) for $B=3.5$ 
from {\it Smoldyn} simulations converge for increasing simulation time. Remaining parameters 
were chosen as in Fig. 4D with $x$ the monomer concentration.
(A-C) Simulation time is increased from $t=1,000$, $5,000$, to $10,000$ as indicated. 
Arrow in (C) points to high state.
(D) Kullback-Leibler divergence between each of the three simulations and {\it Smoldyn} 
simulation for $t=50,000$ as reference distribution (see S1 Text for details).\\
\\
\noindent{\bf S3 Figure:} Comparison of {\it Smoldyn} simulations and Gillespie simulations of 
generalized Schl\"ogl model for increasing diffusion constant $D$ for $B=3.5$ and $t=10,000$. 
Remaining parameters were chosen as in Fig. 4D with $x$ the monomer concentration. 
(A) Gillespie simulation. (B) {\it Smoldyn} simulation for $D=3$ ($X$) and $1$ ($X_2$).
(C) {\it Smoldyn} simulation for faster diffusion with $D=15$ ($X$) and $5$ ($X_2$).
(D) Kullback-Leibler divergence between each of the two {\it Smoldyn} simulations and 
Gillespie simulation as reference distribution (see S1 Text for details).\\
\\
\noindent{\bf S4 Figure:} Distributions from {\it Smoldyn} simulations become increasingly 
similar to results from Gillespie simulations of conventional Schl\"ogl model for 
increasing dimerization rate constants $k_{+3}=k_{-3}$ as indicated, as well as fast diffusion 
(parameters as in S3C Fig.) with enlarged $B=3.7$. (A) Gillespie simulation for $B=3.2$. 
(B) {\it Smoldyn} simulation for $k_{+3}=k_{-3}=1$. (C) {\it Smoldyn} simulation for $k_{+3}=k_{-3}=3$. 
(D) Kullback-Leibler divergence between each of the two {\it Smoldyn} simulations and 
Gillespie simulation as reference distribution (see S1 Text for details).\\
\\
\noindent{\bf S5 Figure:} Drastic reduction of diffusion collapses bistabiliy to predicted 
low monostable state for different $B$ values as indicated (see main text section ``Microscopic 
perspective with diffusion''). Scaling factor multiplies set of diffusion constants, 
$D=3$ ($X$) and $1$ ($X_2$). Remaining parameters as in S3B Fig.\\
\\
\noindent{\bf S6 Figure:} Switching may be triggered by cell-volume changes. 
Image analysis similar to Fig. 8B but with 
total intensity normalized by cell area to provide the intensity density.\\

\end{document}